\documentclass{article}
\usepackage[a4paper,bindingoffset=0.2in,%
             left=0.9in,right=0.9in,top=1in,bottom=1in,%
             footskip=.25in]{geometry}

\usepackage[hang,flushmargin]{footmisc}
\usepackage{graphicx}
\usepackage{amsmath}
\usepackage{amssymb}
\usepackage{booktabs}

\usepackage{booktabs}
\usepackage{graphicx}
\usepackage{multirow}
\usepackage{siunitx,etoolbox}
\usepackage{amssymb}
\usepackage{pifont}
\usepackage{makecell}

\usepackage[inkscapeformat=pdf]{svg}
\usepackage{algorithm}
\usepackage{algpseudocode}
\usepackage{pifont}
\usepackage{caption}
\usepackage{amsfonts} 
\usepackage{url}
\usepackage{cite}
\usepackage{tikz}
\usepackage{comment}
\usepackage{amsmath,amssymb} 
 
\usepackage[accsupp]{axessibility}  

\usepackage[customcolors]{hf-tikz}

\DeclareMathAlphabet{\mcll}{U}{dutchcal}{m}{n}
\usepackage{color}
\usepackage{tcolorbox}
\usepackage{todonotes} 
\newcommand{\bs}{\boldsymbol}
\newcommand{\mbf}{\mathbf}

\newcommand{\mcl}{\mathcal}

\usepackage{xstring}
\def\alphabet{abcdefghijklmnopqrstuvwxyzABCDEFGHIJKLMNOPQRST123456789}
\renewcommand{\vec}[1]{
\IfSubStr{\alphabet}{#1}{
\ensuremath{\mathbf{\MakeLowercase{#1}}}
}{
\ensuremath{\boldsymbol{\MakeLowercase{#1}}}
}
}
\newcommand{\mat}[1]{
\IfSubStr{\alphabet}{#1}{
\ensuremath{\mathbf{\MakeUppercase{#1}}}
}{
\ensuremath{\boldsymbol{\MakeUppercase{#1}}}
}
}

\def\R{\mathbb R}
\def\C{\mathbb C}

\newcommand*{\norm}[1]{\left\|#1\right\|}

\newcommand*{\card}[1]{\left|#1\right|}

\def\defeq{:=}

\usepackage{acronym}
\acrodef{SDE}{Stochastic Differential Equation}
\acrodef{MRI}{Magnetic Resonance Imaging}
\acrodef{DDPM}{Denoising Diffusion Probabilistic Models}
\acrodef{SMLD}{Score Matching with Langevin Dynamics }
\acrodef{NCSN}{Noise Conditional Score Network} 
\acrodef{SSIM}{Structural Similarity Index Measure}
\acrodef{PSNR}{Peak Signal-to-Noise Ratio}

\makeatletter
\def\thanks#1{\protected@xdef\@thanks{\@thanks
        \protect\footnotetext{#1}}}
\makeatother

\usepackage[pagebackref,breaklinks,colorlinks]{hyperref}

\usepackage[capitalize]{cleveref}
\crefname{section}{Sec.}{Secs.}
\Crefname{section}{Section}{Sections}
\Crefname{table}{Table}{Tables}
\crefname{table}{Tab.}{Tabs.}

\def\wacvPaperID{1350} 
\def\confName{WACV}
\def\confYear{2025}
\date{}

\title{MRI Reconstruction with Regularized 3D Diffusion Model (R3DM)}

\author{
  Arya Bangun$^{*1}$ \quad Zhuo Cao$^{*1}$ \quad Alessio Quercia$^{1,3}$ \quad Hanno Scharr$^{1}$ \quad Elisabeth Pfaehler$^{1,2}$
  \thanks{ \hspace{-0.55cm}
  $^{1}$IAS-8, $^{2}$INM-4 Forschungszentrum Jülich, Germany \quad $^{3}$RWTH Aachen University, Germany \newline
\texttt{\{a.bangun, z.cao, a.quercia, h.scharr, e.pfaehler\}@fz-juelich.de}}
}

\begin{document}

\maketitle
\begin{abstract}
\ac{MRI} is a powerful imaging technique widely used for visualizing structures within the human body and in other fields such as plant sciences.
\textcolor{black}{However, there is a demand to develop fast 3D-MRI reconstruction algorithms to show the fine structure of objects from under-sampled acquisition data, i.e., k-space data. This emphasizes the need for efficient solutions that can handle limited input while maintaining high-quality imaging.}
In contrast to previous methods only using 2D, we propose a 3D MRI reconstruction method that leverages a regularized 3D diffusion model combined with optimization method. By incorporating diffusion-based priors, our method improves image quality, reduces noise, and enhances the overall fidelity of 3D MRI reconstructions. We conduct comprehensive experiments analysis on clinical and plant science MRI datasets. To evaluate the algorithm effectiveness for under-sampled k-space data, we also demonstrate its reconstruction performance with several undersampling patterns, as well as with in- and out-of-distribution pre-trained data. In experiments, we show that our method improves upon tested competitors.
\end{abstract}

\section{Introduction} \label{sec:intro}

In order to speed up the acquisition time, MRI instruments acquire sub-sampled k-space data, a technique where only a fraction of the total k-space data points are sampled during the imaging process. Several attempts have been proposed to develop two-dimensional (2D) and three-dimensional (3D) image reconstruction techniques for sub-sampled k-space, as discussed in \cite{muckley2021results, dwork2021fast, chung2022score}. Advancements in 3D MR imaging methods can address the challenges posed by complex anatomical structures of human organs and plant growths. Consequently, the demand for developing 3D MR image reconstruction methods has intensified.
{\let\thefootnote\relax\footnote{{$^*$ Equal contribution}}}
{\let\thefootnote\relax\footnote{{Code: \url{https://jugit.fz-juelich.de/ias-8/r3dm/}}}} 

Currently, most works reconstruct a 3D volumetric image by stacking 2D reconstructions because MR images are acquired slice by slice.
This method doesn't consider the inter-dependency between the slices, thus can lead to inconsistencies and artifacts, as discussed in \cite{chung2023solving, banerjee2021completely, wu20203d}. 
This particularly affects datasets that have equally distributed information and structures with high continuity on all dimensions, such as roots and vessels\cite{schulz2013plant, wu20203d, banerjee2021completely}. 

Before the deep learning-based models, which learn the data-driven prior, the model-based iterative reconstruction method proved its effectiveness in the 3D MRI reconstruction problem \cite{fessler2010model,zhou2023recent}. The problem is formulated as an optimization problem where a data consistency term is applied to ensure fidelity, and a regularisation term, such as the Total Variation (TV) penalty \cite{joshi2009mri} is utilized to provide general prior knowledge of MRI data. Recent developments in generative models, especially the diffusion model, show significant improvement in reconstructing 2D and 3D MR images. This pivots to the idea of combining both data-driven and model-based reconstruction techniques, as presented in \cite{NEURIPS2021_7d6044e9, song2021solving, wen2023conditional, quan2018compressed}. 

One recent works applying the hybrid approach is DiffusionMBIR \cite{chung2023solving}.
The proposed method can be seen as a two-stage process. First, images are generated from a pre-trained 2D diffusion model as in \cite{chung2022score}. Then, optimization methods and regularization techniques are applied on the $z-$axis to produce consistency in the sequence of 3D images. 
In this approach, the consistency of volumetric images highly depends on regularization. One possibility to directly improve the reconstruction of 3D data is leveraging 3D representations of the data-driven prior. As discussed in \cite{wu2021denoising}, 3D representations capture volumetric information more accurately and describe the details of the observed object. This enhancement contributes to a more comprehensive understanding of the underlying data, making the model outperforming its 2D counterpart.

\begin{figure}[htb!]
    \centering
    \includegraphics[width=0.9\textwidth]{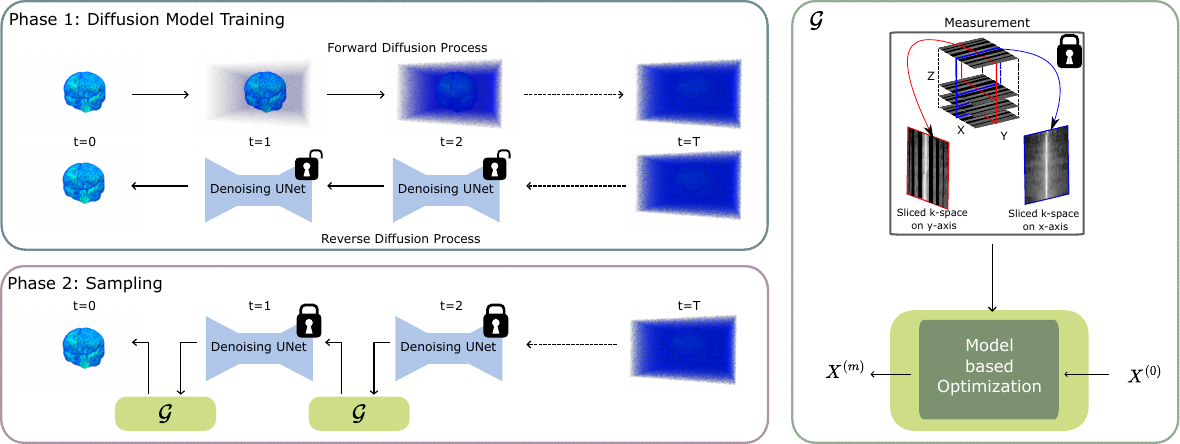}
    \caption{Workflow of the proposed algorithm generating volumetric data from a random distribution and guiding the reconstruction for specific k-space data to have a unique reconstruction. The reverse diffusion process is performed using a pre-trained diffusion model.}
\label{Fig:workflow}
\end{figure}
In this article, we propose a reconstruction method that takes into account the 3D nature of MR images. At the same time, it utilizes a hybrid approach that combines the general prior given by the model-based reconstruction techniques with the data-driven prior from the generative model. The workflow can be seen in Figure \ref{Fig:workflow}.
and our contributions are summarized below:
\begin{itemize}
\item We propose a Regularized 3D Diffusion Model (R3DM), an optimization algorithm to improve 3D MRI diffusion-based reconstruction. 
\item We demonstrate the effectiveness and robustness of our method via extensive experiments that compare its reconstruction quality to that of state-of-the-art methods, with in-distribution and out-of-distribution data.
\item We study the need of 3D representation and that of the model-based method for 3D MRI reconstruction. We demonstrate in ablation studies that both are important for good performance.

\end{itemize}

\section{Related work}\label{sec:background}
In this section, we discuss the state of the art of \ac{MRI} reconstruction from the classical to generative model-based approaches.

\paragraph{Model-based methods}
Conventional methods implemented in current MR scanners for accelerating data acquisition are \textit{e.g.} parallel imaging and compressed sensing \cite{larkman2007parallel, pruessmann2006encoding, lustig2008compressed}. In parallel imaging, multiple receiver coils are used in the acquisition process, where each coil receives an under-sampled amount of k-space data. By incorporating estimated sensitivity maps for each coil, the information of these coils can then be merged to reconstruct one final image. For this purpose, algorithms such as SENSE and  GRAPPA are used \cite{pruessmann1999sense, griswold2002generalized}, where the process to fill the missing information either in image space or k-space is performed. While parallel imaging can accelerate the acquisition time by a factor of 2 or 3, the use of compressed sensing can accelerate the acquisition time by an even higher factor of about 4-8. Compressed sensing uses prior knowledge about the MR signal to iteratively minimize data fidelity and a regularization term and reconstruct the under-sampled k-space data to a high-quality image. With this general prior knowledge, these approaches can be applied to any MR image data. However, to further improve their performance, advanced prior knowledge is needed.

\paragraph{Data-driven methods}
As the aforementioned methods have the limitations of \textit{e.g.} defining the most suitable prior, the use of Deep Learning methods for MR image reconstruction becomes more and more popular \cite{terpstra2020deep, dar2020transfer,aggarwal2018modl,zhu2018image}. Hereby, traditional Convolutional Neural Networks (CNNs) with different architectures have demonstrated promising results such as \textit{e.g.} the Automap framework \cite{zhu2018image} or other deep networks especially designed for MR image reconstruction \cite{eo2018kiki}. Additionally, the effort to accelerate acquisition time by using deep learning has been discussed in \cite{zbontar2018fastmri}, where the challenge is to incorporate a machine learning-based approach to reduce measurement time in the MRI system. Several methods such as fastMRI U-Net, and MoDL\cite{aggarwal2018modl,zbontar2018fastmri, sriram2020end} outperform conventional methods for the MRI reconstruction for undersampling measurements. While these models with learned prior knowledge perform better than the conventional methods, they only work for in-distribution MR images on which the model is trained. Given the high data requirements of deep learning-based methods, training such models requires an effort to collect high-quality data, which is not always possible.

The development of deep learning-based MR image reconstruction is becoming prevalent due to the development of generative models, such as Generative Adversarial Networks (GAN), Normalizing Flows, Denoising Score Matching, and Diffusion Models \cite{mardani2018deep, yang2020pocs,zhang2018multi, song2021solving, chung2022score, wen2023conditional}, where it is demonstrated that those approaches can generate high-quality images from under-sampled k-space data. In those approaches, the model learns the data distribution of the specific MR image, e.g., knee, brain. In the inference process, joint reconstructions are proposed by incorporating under-sampled k-space data for a unique reconstruction. Moreover, the Diffusion Model is robust to out-of-distribution data on which it has not been trained \cite{chung2022score, NEURIPS2021_7d6044e9}, which makes it more data efficient than other deep learning-based models.

In this article, we propose a new hybrid method that combines model-based and data-driven methods.

\section{Methods and Algorithms}\label{sec:methods}
We first briefly review the \ac{MRI} reconstruction problem and diffusion model in sections \ref{sec:mri_recon} and \ref{sec:diffusion_model}, respectively. In \ref{sec:proposed_method} we present the proposed method.
The notations used throughout this article are summarized below.

\paragraph{Notations} \label{sec:notations}
Vectors  $\mathbf{x} \in \C^L$ and matrices are written in bold small-cap and bold big-cap letter $\mathbf{A} \in \C^{K \times L}$, respectively. 
Volumetric data is written as bold italic letter $\textbf{\emph{A}} \in \C^{S \times K \times L}$ or indexing matrix $\mbf{A}_s \in \C^{K \times L} \,\text{for} \, s \in [S]$. Here the $\C$ and $\R$ are complex and real fields. 
The set of integers is written as $[N] := \{1,2,\hdots,N\}$. We denote the two-dimensional and one-dimensional Fourier transform by $\mathcal{F}_{2D}$ and $\mathcal{F}_{1D}$, respectively.
For both matrices and vectors, the notation $\circ$ is used to represent element-wise or Hadamard product. 
For a matrix $\mathbf{X} \in \C^{K \times L}$, the Frobenius norm is denoted by $\norm{\mbf X}_F \defeq \sqrt{\sum_{k = 1}^K \sum_{\ell = 1}^L \card{x_{k\ell}}^2} $. 

\subsection{3D \ac{MRI} Reconstruction} \label{sec:mri_recon}
We first present the conventional 2D reconstruction for full sample k-space data and develop the model for under-sampled measurement. Afterward, we will discuss the challenge of 3D MRI reconstruction.
\paragraph{Full and Under-sampled Reconstruction.}
The relation between fully sampled k-space and the image space, i.e., image of interest,  in \ac{MRI} is given by the 
Fourier transform \cite{slichter2013principles, lin2021artificial, knoll2020deep}. Suppose we acquire $S$ single-coil slices of a fully sampled k-space image $\mbf Y_s \in \C^{N_y \times N_x}$ for $s \in [S]$.   Hence, the direct reconstruction in image space can be written as
\begin{equation}
    \mbf{X}_s = \mcl{F}^{-1}_{2D}\left( \mbf{Y}_s \right) \quad \text{for} \quad s \in [S].
\end{equation}

Accelerating \ac{MRI} acquisition is achieved by reducing the number of scan lines during k-space acquisition. As a result, under-sampled k-space is recorded with fewer data points.  We can write this as
\begin{equation}
    \label{eq:forward_model}
    \hat{\mbf{X}}_s =\mcl{F}^{-1}_{2D}\left(  \mbf{M} \circ {\mbf{Y}}_s  \right) \quad \text{for} \quad s \in [S],
\end{equation}
where the matrix $\hat{\mbf{Y}}_s =  \mbf{M} \circ {\mbf{Y}}_s $ and $\hat{\mbf{X}}_s $ are the noisy under-sampled k-space and the reconstruction, called zero-filled image for all slices, respectively. Applying a two-dimensional inverse Fourier transform directly on under-sampled k-space yields low-quality reconstructed images 
as it results in aliasing and ambiguities.

\paragraph{Challenges in 3D reconstruction.} \label{sec:Challenge3D}
It should be straightforward to construct the 3D image by stacking slice-by-slice reconstruction. However, as discussed in \cite{chung2023solving}, the under-sampled k-space could yield inconsistencies in reconstruction for the $z-$ axis direction. Additionally, reconstruction for the data that have sparse structures in all directions, such as roots and angiography, stacking slice-by-slice reconstruction can also yield non-smoothness in all possible directions. One way to address this problem is by adding constraints in the reconstruction algorithm such that the continuity on $z-y$, $x-z$, as well as $x-y$ can be pertained. Here, we will use the knowledge of projection of 3D data on the image space and its relation to the k-space by using the relation between 3D k-space and projection of 3D image space, as described in Appendix \ref{sec:FourierSlice}, combined with 3D diffusion model. The visualization can be seen in Figure \ref{Fig:projection_image}

\subsection{Diffusion Model} \label{sec:diffusion_model}
The key idea in diffusion models is to learn the training data distribution, i.e., image, such that we can generate new images corresponding to the learned distribution.
\paragraph{Training.} In \ac{DDPM} \cite{ho2020denoising, ho2022video}, estimating data distribution can be performed by perturbing the training data with a sequence of positive noise scale $0 < \beta_1, \beta_2, \hdots, \beta_T < 1$ as  follows 
$$
\mbf{x}_t = \sqrt{1 - \beta_t} \mbf{x}_{t-1} + \sqrt{\beta_t} \mbf{z}_{t-1},\,\text{for}\, t \in \{1,2, \hdots, T \}
$$
where $\mbf{z} \sim \mcl{N}\left(\bs{0}, \mbf{I} \right)$.  
In this case, the original data $\mbf{x}_0 \sim p_{\text{data}}\left(\mbf x\right)$ is perturbed by a random Gaussian vector with increased variance for each time index. Hence, for each time step of perturbation, the neural network model $\mcll{s}_{\bs \theta}$, for instance, the U-Net architecture, is trained to approximate the conditional distribution from perturbed data.

\paragraph{Sampling.} The pre-trained neural network model $\mcll{s}_{\bs \theta^*} $ is used later to remove the perturbation aspect in order to generate an image similar to the training data distribution. The sampling\footnote{In this article the term sampling should not be confused with undersampling in MRI definition. Here sampling in the generative model describes generating data from a certain probability distribution.} procedure is usually performed, for instance with the Euler-Maruyama scheme, as procedure, as follows:
\begin{equation}
\mbf{x}_{t-1} = \frac{1}{\sqrt{1 - \beta_t}} \left(\mbf{x}_t + \beta_t \mcll{s}_{\bs \theta^*} \left(\mbf{x}_t,t\right) \right) + \sqrt{\beta_t} \mbf{z}_t  \,\,\,\text{for}\, t \in \{T,T-1,\hdots,1 \},
\label{eq:sampling}
\end{equation}
It should be noted that, the index is reversed because we perform reverse process in the sampling mechanism. For the complete derivation, we refer the interested reader to the literature and references therein \cite{song2020score,ho2020denoising, ho2022video}.

\paragraph{Diffusion Model for MRI Reconstruction} 
Sampling from a pre-trained diffusion model generates MR images randomly. Guidance is needed to reconstruct a high-fidelity image conditioned on the measurement. Previous works \cite{chung2022come, chung2022improving} proposed to solve the reconstruction, or inverse problem in general, using conditional diffusion models. The idea is to alternatively update the sampling and optimization steps so that the transformation of the generated images asymptotically approach to the measurements. We apply the same procedure in our proposed method. 

\subsection{Proposed Method} \label{sec:proposed_method}

 In this article, we propose an optimization method for 3D reconstruction by leveraging information about the 2D-projected image and its relation to the measurement data (k-space). The relation is from the nature of Fourier transform on 3D image space \cite{bracewell1990numerical, bracewell1956strip}. The algorithm allows us to reconstruct 3D MRI data with enhanced continuity on the $z-$axis using a pre-trained Diffusion Model, i.e. the algorithm is only applied during the sampling process.

\paragraph{3D Representation of data-driven prior}
As mentioned in Section \ref{sec:intro}, a 3D representation that captures the volumetric information is needed for an accurate 3D reconstruction. 
However, training and sampling a 3D Diffusion Model is in general computationally expensive. As highlighted in \cite{voleti2022mcvd}, the 2D U-Net with cross-attention on the 3rd dimension is more efficient than the 3D U-Net version for video data generation. As MR images are taken sequentially, the 3rd dimension of MR images can be regarded as temporal dimension. We called this architecture the 2D+A architecture hereafter. In order to directly reconstruct a 3D volumetric image from a given sub-sampled k-space, the 2D+A diffusion model should be combined with an algorithm that directly processes 3D volumetric data. 
In Section \ref{Sec:Ablation}, we compare the performance with and without cross attention to demonstrate its importance.

\paragraph{Optimization problem.}
To incorporate the measurement process in the image generation in MRI reconstruction, a certain relation between k-space data and the target image space is developed. In this section, we discuss the connection of 3D k-space data and 3D image space.

Suppose we have under-sampled 3D k-space data given as $\mbf{\hat Y}_s \in \C^{N_y \times N_x}$ for all depth slices $s \in [S]$. The zero frequency of k-space at $x-$axis\footnote{Since the matrix represents the image per slice, the zero frequency on $x-$ and $y-$ axis is given in the middle of row or column space}, i.e., $k_x = 0$, can be written as $  {\mbf{\hat y}}_s^{k_y} = \mbf{\hat Y}_s \vert_{\left(k_x = 0\right)} \in \C^{N_y}$, here we use superscript $k_y$ to inform that this vector is obtained for $k_x = 0$. Additionally, the projection of the 3D image space on the $y-$ axis is given by summing in $x-$direction, i.e.
${x}_{s,i} = \mcl{P}_{y}\left(\mbf{X}_s\right)|_i =  \sum_{j = 1}^{N_x} x_{s,i,j} \in \C$.  The relation between the under-sampled k-space slice and the projection image is given by 
\begin{equation}
     \hat{\mbf{y}}_s^{k_y} = \mbf{m}^{k_y} \circ \mcl{F}_{1D}\left( \mbf{x}_s\right) \in \C^{N_y} \quad  \text{for}\quad s \in [S],
\end{equation}
where $\mcl{F}_{1D}$ is the one-dimensional Fourier transform and $ \mbf{m}^{k_y}= \mbf{M} \vert_{\left(k_x = 0\right)} \in \C^{N_y}$ is the slice of the under-sampling operator at $k_x = 0$. The same concept can be adopted to obtain the projected image space on the $x$-axis from the projecting function $\mcl{P}_{x}$. The complete derivation is presented in Appendix \ref{sec:FourierSlice}. \Cref{Fig:projection_image} visualizes the relationship of projected volume data and image space and its relation to the Fourier transform.
\begin{figure}[tb]
\centering
    \includegraphics[width=0.7\textwidth]{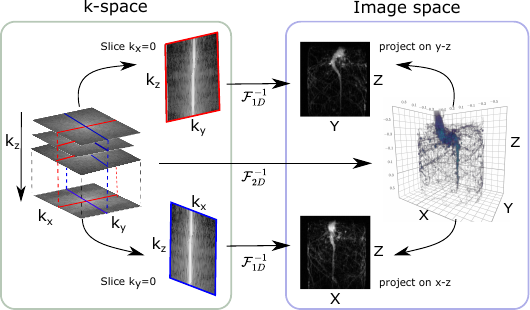}
    \caption{Relation between the projected image and a slice of k-space from plant roots data}
\label{Fig:projection_image}
\end{figure}
From this information, we can form an optimization problem to reconstruct three-dimensional images from under-sampled k-space, as follows:
\begin{equation}
\small
\underbrace{\begin{aligned}
\label{eq:opt}
 \textbf{\emph X$^*$} = &\underset{\textbf{\emph X} \in \C^{S \times N \times N}}{\text{arg min}}
& & 
\sum_{s = 1}^S \norm{\hat{\mbf{Y}}_s - \mbf{M} \circ \left( \mcl{F}_{2D}\left( \mbf{X}_s \right)\right) }_F^2
\tikzmarkend{a-1}
+ \mcl{R}\left(\textbf{\emph X} \right) \\
& \text{subject to}
& & \hat{\mbf{y}}_s^{k_y} = \mbf{m}^{k_y} \circ \mcl{F}_{1D}\left( \mcl{P}_{y} \left(\mbf{X}_s\right)\right) \quad  \text{for } s \in [S]\\
&&& \hat{\mbf{y}}_s^{k_x} =  \mbf{m}^{k_x} \circ \mcl{F}_{1D}\left( \mcl{P}_{x} \left(\mbf{X}_s\right)\right) 
\end{aligned}}_{\mcl{G}\left(\textbf{\emph X} \right)}
\end{equation}
The regularization functions $\mcl{R}\left(\textbf{\emph X}\right)$ used in this article can be written as
\begin{equation}
\label{eq:reg}
\begin{aligned}
  &\alpha \underbrace{\sum_{s=1}^S\sum_{i=1}^N\sum_{j=1}^N \card{x_{s,i,j}}}_{\text{sparsity}} +\underbrace{\sum_{s=1}^S \sum_{i=1}^N\sum_{j=1}^{N-1}  \card{x_{s,i,j+1} - x_{s,i,j}}^2}_{\text{smoothness}}   &+  \underbrace{\sum_{s=1}^S\sum_{i=1}^{N-1}\sum_{j=1}^{N} \card{x_{s,i + 1,j} - x_{s,i,j}}^2}_{\text{smoothness}} 
\end{aligned}
\end{equation}
In general, the optimization problem $\mcl{G}\left(\textbf{\emph X} \right)$ as in \eqref{eq:opt} describes the objective and the constraint functions to estimate the 3D image space given the relation between its under-sampled 3D k-space and the projection image. This ensures the reconstructed image corresponds to the given measurement. 

 \begin{algorithm}
    \caption{Regularized 3D Diffusion Model (R3DM)}
    \label{algo:DDPM_Sampling_Proximal}
    \begin{algorithmic}[1]
        \State \textbf{Initialization:} 
        \begin{itemize}
            \item Timesteps $T$ and optimization iterations $m$
            \item Volume sampling from prior distribution $ {\textbf{\emph X}_{T}} \in \R^{S \times N \times N} \sim \mcl{N}\left(\bs{0},\mbf{I}\right)$ 
        \end{itemize}
        \For{each reverse iteration $t = T \quad \textbf{to} \quad 1 $ }
            \State Generate $\textbf{\emph Z}  \sim \mcl{N}\left(\bs{0},\mbf{I}\right)$ if $t > 1$ else $\textbf{\emph Z} = \bs{0}$  
            \State Update sample {(\ref{eq:sampling})} $$
            \textbf{\emph X}^{\left(0\right)} = \frac{1}{\sqrt{1 - \beta_t}} \left(\textbf{\emph X}_t + \beta_t \mcll{s}_{\bs \theta^*} \left(\textbf{\emph X}_t,t\right) \right) + \sqrt{\beta_t} \textbf{\emph Z}_t 
            $$
            {\For{each iteration $i = 1 \quad \textbf{to} \quad m $ } 
                \State{\parbox[t]{\dimexpr\linewidth-\algorithmicindent}{%
          Solving the optimization\\ problem in \eqref{eq:opt}\\
          $\textbf{\emph X}^{\left(i\right)} = \mcl{G}\left(\textbf{\emph X}^{\left(i-1\right)}\right)$
        }}
                
            \EndFor
            \State $\textbf{\emph X}_{t-1} = \textbf{\emph X}^{\left(m\right)} $}
        \EndFor\\
        \Return ${\textbf{\emph X}}_0 \in \C^{S \times N \times N}$
    \end{algorithmic}
\end{algorithm}

\paragraph{Regularization} 
In addition, we use regularization for each element of volumetric data in terms of $\ell_1$-norm that enforces sparsity structure, left side constraint in \eqref{eq:reg}, and the approximated total variation norm that enforces smoothness, right side constraint in \eqref{eq:reg}. These terms are used for imposing the result with certain structures, i.e., sparsity or smoothness \cite{parikh2014proximal, rudin1992nonlinear}. The importance of the regularization terms is discussed in Section \ref{Sec:Ablation}.
In practice, the implementation can be performed in terms of alternating minimization combined with proximal methods for the regularization function. Note that the $\ell_1$-norm constraint is non-smooth, the proximal method is used to avoid calculating its derivative. The analytic proximal of $\ell_1$-norm is the well known soft thresholding function  \cite{bolte2014proximal, parikh2014proximal, foucart2013mathematical}, which is also presented in theAppendix \ref{Sect:Prox_Regul}. Hence, the algorithm can be seen as applying a proximal function on the gradient update of the loss function. This is done by applying all constraints in the optimization formula \eqref{eq:opt} as a single loss equation. This process is also called proximal gradient method \cite{bolte2014proximal, parikh2014proximal}. The implementation detail will be presented in Appendix \ref{Sect:Prox_Regul}. 

\paragraph{Regularized 3D Diffusion Model (R3DM) }
The above optimization steps are inserted in the sampling process of a pre-trained Diffusion Model. Therefore, the proposed Regularized 3D Diffusion Model (R3DM), consists of two parts. The Diffusion Model part provides the data-related prior knowledge. The optimization method incorporates the k-space measurement to ensure the fidelity of the generated data. In addition, the regularisation term in \eqref{eq:reg} gives general prior knowledge of the MRI data, i.e. sparsity and smoothness. 
The overall algorithm is described in Algorithm \ref{algo:DDPM_Sampling_Proximal}. The algorithm starts with generating a 3D MR image via the sampling process of \ac{DDPM}, which is guided by the optimization method in \eqref{eq:opt} to incorporate the under-sampled k-space. I.e. the optimization method is guided such that a specific 3D MR image is reconstructed from the k-space measurement. It should be noted that the superscript $(i)$ and subscript index $t$ refers to the iteration index between both optimization update and the \ac{DDPM} sampling. Hence, the algorithm performs \ac{DDPM} sampling and optimization updates alternately.

$$
$$
$$
$$

\vspace{-2.cm}
\section{Experiments}\label{sec:experimental}
\begin{figure*}[htb!]
    \centering
     \includegraphics[width=0.95\textwidth]{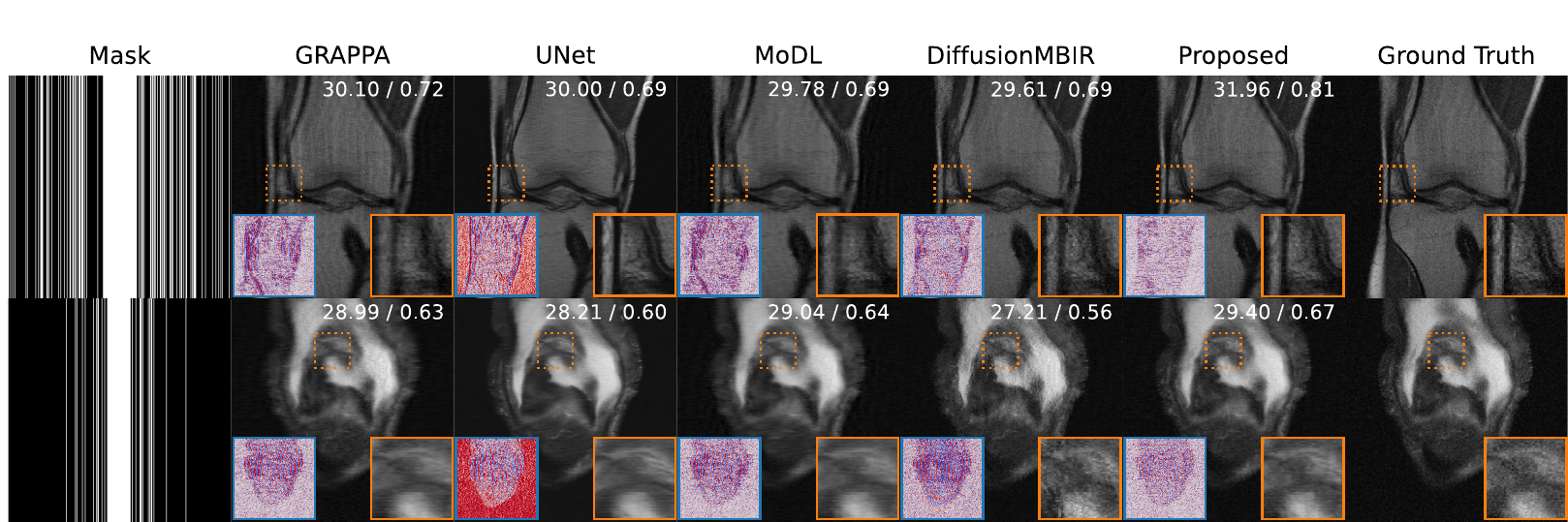}
    \caption{Single slice from the volume reconstruction of file1000758 (top) and file1001862 (bottom) from fastMRI knee data. 
    The numbers on the top right corner represent the PSNR/SSIM of the slice. The subplots on the lower left corner represent the difference map between the reconstruction and ground truth. The color range is between $-0.02$ (bluish) and $0.02$ (reddish). Note that the volumetric ground truth data has been normalized. The subplots on the lower right are a zoomed-in view.}
\label{Fig:compare_knee}
\end{figure*}
\begin{table*}[tb]
\newrobustcmd{\rankfirst}{\fontseries{b}\selectfont}
\centering
\caption{Mean of SSIM and PSNR from fastMRI knee data. The \textbf{bold} and \underline{underline} represent the best and second best result.}
\label{tab:result_knee}
\resizebox{1.\textwidth}{!}{
    \begin{tabular}{@{} c l 
                    *{4}{S[table-format=2.4,detect-weight,mode=text]} 
                    *{4}{S[table-format=3.3,detect-weight,mode=text]} 
                    @{} }
    \toprule
    \multicolumn{2}{c}{}   & \multicolumn{4}{c}{SSIM $(\uparrow)$}      & \multicolumn{4}{c}{PSNR ($\uparrow$)}      \\
    \cmidrule[0.5pt](rl){3-6}
    \cmidrule[0.5pt](rl){7-10}
    Mask Types      &     Methods     & {3D}    & Axial    & Sagittal    & Coronal    & {3D}    & Axial    & Sagittal    & Coronal    \\   \hline
    \multirow{6}{*}{\makecell{Uniform \\ (2x, 0.15)}}   
                              
                              & GRAPPA\cite{griswold2002generalized}  & \underline{0.868} & \underline{0.733} & \underline{0.783} & \underline{0.805} & \underline{31.33} & \underline{29.15} & \underline{26.27} & \underline{27.53} \\
                              & U-Net\cite{unetmri2017}                & 0.839 & 0.687 & 0.743 & 0.770 & 30.36 & 28.01 & 25.24 & 26.46 \\
                              & MoDL \cite{aggarwal2018modl}   & 0.867 & 0.721 &  0.792 & 0.807 & 31.12 & 28.92 & 26.12 & 27.28 \\
                              & DiffusionMBIR\cite{chung2023solving}  & 0.850 & 0.700 & 0.764 & 0.783 & 30.53 & 28.23 & 25.48 & 26.67 \\
                              & Proposed                              & \rankfirst 0.909 & \rankfirst 0.818 & \rankfirst 0.859 & \rankfirst 0.870 & \rankfirst 32.89 & \rankfirst 30.63 &  \rankfirst 27.80 & \rankfirst 29.02 \\
                      
                              \hline
    \multirow{6}{*}{\makecell{Gaussian  \\ (8x, 0.08)}} 
                              & GRAPPA          & \underline{0.794} & \underline{0.572} & \underline{0.672} & \underline{0.698} & {28.81} & {26.68} & {23.88} & {25.05} \\
                              & U-Net            & 0.789 & 0.561 & 0.661 & 0.692 & \underline{29.25} & \underline{26.93} & \underline{24.21} & \underline{25.37} \\
                              & MoDL     & 0.795 & 0.566 & 0.678 & 0.701 & 29.01 & 26.81 & 24.07 & 25.24 \\
                             & DiffusionMBIR   & 0.761 & 0.514 & 0.631 & 0.656 & 28.08 & 25.78 & 23.13 & 24.27 \\
                              & Proposed        & \rankfirst 0.823 & \rankfirst{0.626} & \rankfirst 0.722 & \rankfirst 0.743 & \rankfirst{29.83} &  \rankfirst{27.61}  &  \rankfirst{24.84} & \rankfirst{26.00}\\ 
                              \hline
    \end{tabular}%
}
\end{table*}

In this section, we discuss the evaluation of proposed algorithms in terms of several metric, such as structural similarity index (SSIM) and peak signal-to-noise ratio (PSNR).   

\subsection{Experiment setup} 
\paragraph{Datasets}\label{sec:data}
In this paper, we use one dataset consisting of MR images of the knee from the fastMRI dataset  \cite{zbontar2018fastmri} for training, and three datasets for testing our approach, namely MR images of the knee from the fastMRI dataset  \cite{zbontar2018fastmri}, the BRATS brain \cite{menze2014multimodal, bakas2017advancing, bakas2018identifying}, and plant roots \cite{van2016quantitative,pflugfelder2017non}. For the knee data, we use the central slices to avoid noise-only images. Therefore the input data dimension is $N \times N \times S = 320 \times 320 \times 16$ with voxel size of 0.5 mm $\times$ 0.5 mm $\times$ 3 mm. All $973$ training volumetric knee data from fastMRI are used for training.
The brain dataset BRATS has a dimension of $ 240 \times 240 \times 155$. The voxel size of each image is 1 mm$^3$.
For plant roots data the full k-space data are acquired with a single coil MR scanner. The k-space data as well as the reconstructed images yield the image dimension $192 \times 192 \times 100$ and voxel size 0.5 mm  $\times$ 0.5 mm $\times$ 0.99 mm. 
To evaluate in-distribution reconstruction, we use $30$ randomly chosen volumetric data provided in the validation set of fastMRI knee data.
For out-of-distribution, we evaluate the algorithm with $30$ and $25$ randomly chosen volumetric data for BRATS and plant roots, respectively.  
During the evaluation of BRATS data, the noise-only slices are removed. 
 
\paragraph{Diffusion Model Training Procedure}\label{sec:model}
The experiments conducted below utilize the model that was trained on the fastMRI knee dataset \cite{zbontar2018fastmri}. We trained the model for $70,000$ iterations using $\ell_1$ loss and a learning rate of $10^{-4}$. Due to the large size of the 3D data, the batch size is set to $1$ while a gradient accumulation technique is applied to ensure training stability. In general, we follow the parameter setting given in \cite{ho2020denoising} for training a U-Net to estimate the data distribution. 
\vspace{-0.3cm}
\paragraph{Description of the comparative methods}
We perform a comparison with several existing algorithms for \ac{MRI} reconstruction, namely GRAPPA \cite{griswold2002generalized}, fastMRI U-Net \cite{unetmri2017}, MoDL \cite{aggarwal2018modl}, and  DiffusionMBIR \cite{chung2023solving}. For each method, we perform reconstruction with the same under-sampling operator, i.e., masking. In contrast to all other methods, GRAPPA is a classical reconstruction algorithm that is applied directly for image reconstruction of under-sampled k-space data. We report method details in the Appendix \ref{App:Method_details}.
\begin{table*}[tb]
\newrobustcmd{\rankfirst}{\fontseries{b}\selectfont}
\centering
\caption{Mean of SSIM and PSNR from BRATS and plant roots data. The \textbf{bold} and \underline{underline} represent the best and second-best results. }
\label{tab:result_roots}
\resizebox{1.\textwidth}{!}{
    \begin{tabular}{@{} c c l 
                    *{4}{S[table-format=2.4,detect-weight,mode=text]} 
                    *{4}{S[table-format=3.3,detect-weight,mode=text]} 
                    @{} }
    \toprule
    \multicolumn{3}{c}{}   & \multicolumn{4}{c}{SSIM $(\uparrow)$}      & \multicolumn{4}{c}{PSNR ($\uparrow$)}      \\
    \cmidrule[0.5pt](rl){4-7}
    \cmidrule[0.5pt](rl){8-11}
    Dataset & Mask Types      &     Methods     & {3D}    & Axial    & Sagittal    & Coronal    & {3D}    & Axial    & Sagittal    & Coronal    \\   \hline
         \multirow{12}{*}{Roots} 
        & \multirow{6}{*}{\makecell{Uniform \\ (2x, 0.15)}}   
              & GRAPPA\cite{griswold2002generalized}   & \underline{0.966} & \underline{0.878} & \underline{0.795} & \underline{0.797} & \underline{44.40} & \underline{37.39} & \underline{32.21} & \underline{32.16} \\
        &      & U-Net\cite{unetmri2017}                & 0.853 & 0.731 &  0.630 & 0.629 &  37.71 & 30.81 & 26.01 & 25.93 \\
        &      & MoDL \cite{aggarwal2018modl}   & 0.963 & 0.872 & 0.778 & 0.789 & 43.47 & 36.52 & 31.54 & 31.63 \\
        &      & DiffusionMBIR\cite{chung2023solving}  &0.908  &0.666 & 0.504 & 0.494 & 40.18 & 32.93 & 27.79 & 27.55 \\
        
        &      & Proposed                              & \rankfirst 0.975 & \rankfirst 0.902 & \rankfirst 0.835 & \rankfirst 0.830 & \rankfirst 46.92 & \rankfirst 39.68 & \rankfirst 34.54 & \rankfirst 34.30 \\
        \cmidrule[0.5pt](rl){2-11}
        & \multirow{6}{*}{\makecell{Gaussian  \\ (8x, 0.08)}} & GRAPPA          & \underline{0.945} & \underline{0.802} & \underline{0.666} & \underline{0.660} & \underline{41.83} & \underline{34.88} & \underline{29.87} & \underline{29.78}\\
        &      & U-Net            & 0.885 & 0.725 & 0.580 & {0.573} & {38.32} &  {31.59} & {26.94} & {26.96} \\
        &      & MoDL  & 0.941 & 0.797 & 0.656 & 0.653 & 41.09 & 34.24  & 29.41 & 29.52  \\
        &      & DiffusionMBIR   & 0.876 & 0.588 & 0.418 & 0.409 & 38.58 & 31.34 & 26.21 & 25.97 \\
        
        &      & Proposed        & \rankfirst 0.954 & \rankfirst 0.823 & \rankfirst 0.698 & \rankfirst 0.687 & \rankfirst 44.11 & \rankfirst 36.93 & \rankfirst 31.82 & \rankfirst 31.59\\ 
   
    \toprule
 \multirow{12}{*}{BRATS} 
        & \multirow{6}{*}{\makecell{Uniform \\ (2x, 0.15)}}   
        & GRAPPA & \underline{0.950} & \underline{0.927} & \underline{0.891} & \underline{0.892} & \underline{37.70} & 35.60 & \underline{31.48} & \underline{33.58} \\
        &    & U-Net               &  0.806 & 0.675 & 0.691 & 0.653 & 36.69 & 33.23 & 30.93 & 31.74  \\
        &      & MoDL   & 0.950  & 0.927 & 0.887 & 0.883 & 37.84 & 35.21 & 31.81 & 32.70 \\
        &    & DiffusionMBIR  & 0.947 & 0.867 & 0.859 & 0.858 & \rankfirst{41.12} & \underline{37.31} & \rankfirst{36.17} & \rankfirst{36.36} \\
        &    & Proposed                              & \rankfirst 0.971 &\rankfirst 0.958 & \rankfirst0.932 & \rankfirst 0.938  & 39.43 & \rankfirst{38.18} & 33.29 & 36.50  \\
        \cmidrule[0.5pt](rl){2-11}
        & \multirow{6}{*}{\makecell{Gaussian  \\ (8x, 0.08)}} 
        & GRAPPA          & 0.894 & \underline{0.848} & \underline{0.834} & \underline{0.779} & 33.97 & 31.40 & 27.80 & 29.48 \\
        &    & U-Net           & 0.742 & 0.610 & 0.654 & 0.545 & \underline{34.69} & 31.30 & \underline{28.98} & \underline{29.68} \\
   
          &      & MoDL   & 0.872 & 0.821 & 0.817 & 0.735  & 33.93  & 31.01 & 27.95 & 28.88  \\
        &    & DiffusionMBIR   & \underline{0.909} & 0.796 & 0.780 & \underline{0.779} & \rankfirst{37.21} & \rankfirst{33.54} & \rankfirst{31.79} & \rankfirst{32.27} \\
        
        &    & Proposed        & \rankfirst{0.932} & \rankfirst{0.908} & \rankfirst{0.856} & \rankfirst{0.863} &  {33.72} & \underline{31.43} & 27.61 & 29.85 \\
        \hline
    \end{tabular}%
}
\end{table*}  
\subsection{Results}
Volumetric reconstruction is performed with under-sampled k-space data from the specified under-sampling operator. For each algorithm, we calculate the image quality metrics SSIM and PSNR \cite{gourdeau2022proper, wang2004image, hore2010image}. 
We report the average of those metrics for the testing datasets. For our optimization method $\mcl{G}\left(\textbf{\emph X}\right)$ in \eqref{eq:opt}, we use the learning rate $\eta=0.01$, number of iteration $m=10$, scaling for $\ell_1$-norm $\alpha=0.02$ and approximated total variation $tv=1$, as a result of a hyper-parameter grid search, reported in Appendix \ref{App:Hyperparameters}. In what follows, we will discuss the reconstruction performance for in-distribution and out-of-distribution data.
\paragraph{In Distribution}
In-distribution reconstruction measures the performance of the network model on data from the same distribution as the training data. We perform an evaluation of in-distribution reconstruction by using pre-trained fastMRI knee data for fastMRI U-Net, MoDL, DiffusionMBIR, and our method. In contrast to this, GRAPPA works directly without pre-training.  Table \ref{tab:result_knee} shows the mean SSIM and PSNR values for all methods applied on several under-sampling measurements.  
Apart from directly evaluating the SSIM and PSNR per volume, we also evaluate those metrics for all 2D slices of x, y, and z axes. 

We observe that in Uniform masking, the proposed algorithm outperforms other methods in terms of SSIM and PSNR. Given the $8\times$ acceleration factor applied in Gaussian masking, there is a decrease in both overall SSIM and PSNR when compared to Uniform masking with a $2\times$ acceleration factor due to the reduced amount of information available in the k-space data for the reconstruction process. For SSIM evaluation the proposed algorithm still performs better than other algorithms. A single slice taken from the 3D reconstruction of fastMRI knee data for different each mask type is presented in \Cref{Fig:compare_knee}. Overall, all methods yield visual similar results, however with clearly noticeable differences on close inspection. For $2\times$ as well as for $8\times$ acceleration we observe that DiffusionMBIR reconstructs somewhat noisy-looking images, while fastMRI U-Net yields crisp-looking results with the highest contrast. However, as the metric scores support, the high contrast surpasses the contrast of the ground truth and this may indicate mild hallucination of structures. In addition, it smooths low-contrast regions generating a mildly cartoonish impression. GRAPPA produces blurrier images, especially in the $8\times$ case. Quantitatively, the proposed method is closest to the ground truth.
\begin{table*}[!ht]
\newrobustcmd{\rankfirst}{\fontseries{b}\selectfont}
\centering
\caption{Mean of SSIM and PSNR for 2D + A and stacking 2D Architecture to generate 3D knee images. The \textbf{bold} represents the best result. The regularization terms are not activated.}
\label{tab:ablation_2D_2D_T}
    \begin{tabular}{@{} c l 
                    *{4}{S[table-format=2.4,detect-weight,mode=text]} 
                    *{4}{S[table-format=3.3,detect-weight,mode=text]} 
                    @{} }
    \toprule
    \multicolumn{2}{c}{}   & \multicolumn{4}{c}{SSIM $(\uparrow)$}      & \multicolumn{4}{c}{PSNR ($\uparrow$)}      \\
    \cmidrule[0.5pt](rl){3-6}
    \cmidrule[0.5pt](rl){7-10}
    Mask Types      &     Methods     & {3D}    & Axial    & Sagittal    & Coronal    & {3D}    & Axial    & Sagittal    & Coronal    \\   \hline
    \multirow{2}{*}{\makecell{Uniform \\ (2x, 0.15)}}   
                              
                              & 2D + A \cite{ho2022video}  & \rankfirst{0.815} & \rankfirst{0.648} & \rankfirst{0.742} & \rankfirst{0.756} & \rankfirst{28.41} & \rankfirst{26.61} & \rankfirst{23.44} & \rankfirst{24.70} \\
                              & 2D \cite{ho2020denoising}                     & {0.769} & {0.579} & {0.684} & {0.698} & {26.88} & {24.57} & {21.93} & {23.29}  \\
                              \hline
    \multirow{2}{*}{\makecell{Gaussian  \\ (8x, 0.08)}} 
                              & 2D + A          & \rankfirst{0.673} & \rankfirst{0.404} & \rankfirst{0.567} & \rankfirst{0.584} & \rankfirst{25.46} & \rankfirst{23.44} & \rankfirst{20.54} & \rankfirst{21.65} \\
                              & 2D           & {0.558} & {0.304} & {0.454} & {0.464} & {22.79} & {20.38} & {17.82} & {19.02} \\ 
                              \hline
    \end{tabular}
\end{table*}

\paragraph{Out of Distribution}
An out-of-distribution (OOD) evaluation shows the robustness of algorithms for the reconstruction task. As discussed in Section \ref{sec:intro}, general prior knowledge can be applied to any MRI data, however, with limited representativeness. While deep learning-based methods usually only work for in-distribution data, the hybrid Diffusion Model has been shown to be robust to OOD data where it has not been trained \cite{chung2022score, NEURIPS2021_7d6044e9}. The proposed algorithm utilizes both conventional optimization methods and diffusion models and is therefore expected to achieve good performance in OOD evaluation.

We here measure the performance of all methods applied to OOD data, i.e., to reconstruct images that are different from the training images. Table \ref{tab:result_roots} presents the evaluation of average \ac{SSIM} and \ac{PSNR} for both plant roots and BRATS data. In BRATS case, the fastMRI U-Net performs the worst compared to all other algorithms. Meanwhile, the Diffusion MBIR struggles to perform for plant roots data. 
%
%
The proposed method effectively enhances the reconstruction, specifically improving the structure and feature properties of the images. Since each slice in the roots data only represents the sparse distribution of a root, we perform the maximum intensity projection on each axis to visualize the fine roots structure.  When assessing PSNR for BRATS data,DiffusionMBIR mostly outperforms other methods.  This observation suggests that the Diffusion MBIR primarily contributes to reducing the absolute error between pixels and the ground truth than features and structures of images. We report qualitative results for the BRATS and plant roots data  Appendix \ref{App:Qualitative}.
\begin{figure}[!ht] 
\centering
\includegraphics[width=0.6\textwidth]{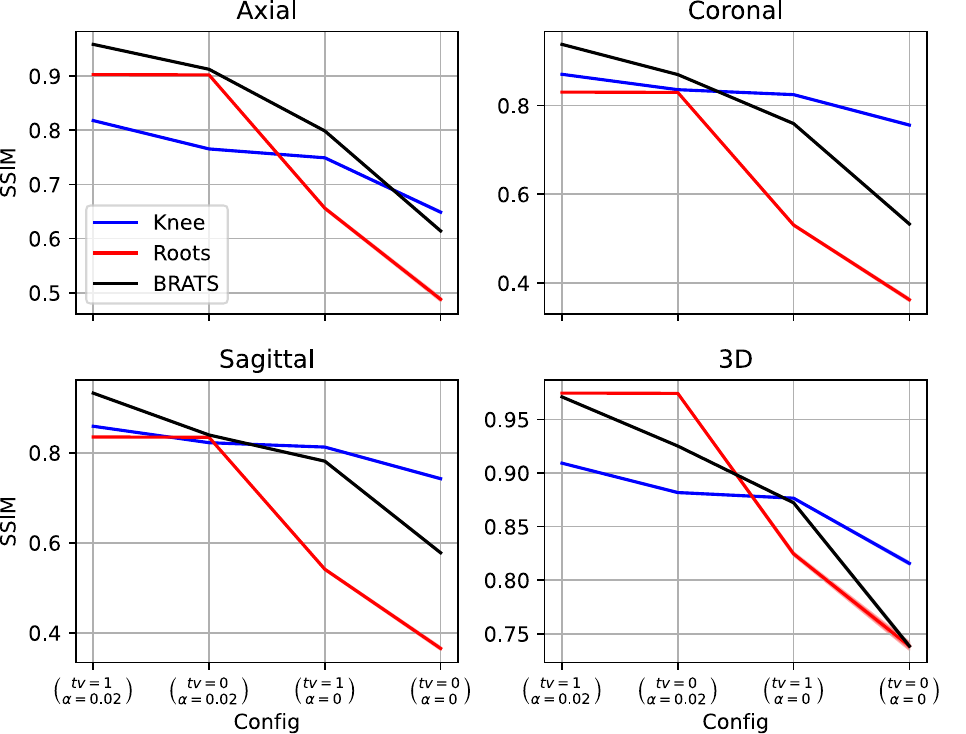}
\caption{Ablation study to investigate the effect of regularization for reconstruction in terms of SSIM $(\uparrow)$ metric for uniform mask. The optimization iteration $m=10$ and the learning rate $0.01$.}
\label{Fig:ablation_uniform}
\end{figure}
\vspace{-0.3cm}
\section{Ablation studies} 
\label{Sec:Ablation}
We perform ablation studies to highlight the effects of regularization on the optimization method and to compare the impact of the 2D + A architecture versus the stacking 2D architecture in generating 3D images
\subsection{Importance of regularization}
In  \Cref{Fig:ablation_uniform}, we demonstrate that both regularization terms are essential for improving reconstruction based on the SSIM metric across all datasets. In the roots dataset, dominated by sparse regions, the $\ell_1$-norm significantly influences performance more than the approximated total variation ($tv$).
Conversely, in the knee dataset, both the approximated total variation and $\ell_1$-norm exhibit similar effects on performance. The result for the Gaussian mask is presented in \Cref{App:ab_gaussian}.
\subsection{Importance of 3D representation} While the 2D + A architecture directly generates 3D volumetric images, the 2D architecture constructs 3D volumetric images by stacking the 2D image slice by slice for both Uniform and Gaussian masks. It can be seen that, as discussed in Section \ref{sec:Challenge3D}, stacking 2D images to generate a 3D image leads to inconsistency as shown in Table \ref{tab:ablation_2D_2D_T}, yielding low SSIM metric for all combination axes. 
\vspace{0.5cm}
\section{Conclusion}\label{sec:conclusion}
In this study, we present a novel 3D reconstruction approach by integrating 2D + A diffusion model and proposed anoptimization method as a guiding constraint for refining the diffusion model's sampling process. Our proposed method enables a pre-trained 3D diffusion model to faithfully reconstruct MR images from both in- and out-of-distribution input data. This capability has been validated both quantitatively and qualitatively through comprehensive comparisons with existing methods. For fastMRI Knee and plant roots datasets, the proposed method outperforms all other tested methods.

Moving forward, we plan to investigate various diffusion model architectures to further enhance the performance and capabilities of the proposed method for a broader range of applications. Moreover, we aspire to augment the quantity and diversity of data utilized by the model, thereby enriching its generalization capabilities. Additionally, we are keen on enabling parallelization techniques to facilitate real-time inference, thus advancing the practical applicability of the models in real-world scenarios.

\vspace{0.2cm}
\section*{Acknowledgment}
This work was supported by the President’s Initiative and Networking Funds of the Helmholtz Association of German Research Centres [Grant HighLine ZT-I-PF-4-042]. The authors gratefully acknowledge computing time on the supercomputer JURECA\cite{thornig2021jureca} at Forschungszentrum Jülich under grant no. \texttt{delia-mp}.
\newpage
{\small
\bibliographystyle{ieee_fullname}
\bibliography{main}
}

 
\newpage 
\appendix
\section{Supplementary Materials}
\section{Methods details}
\label{App:Method_details}
As GRAPPA is frequently used in current MR scanners, it is necessary to include it in the comparison. We perform grid search of GRAPPA parameters that yield the highest possible SSIM and PSNR, namely kernel size $7 \times 7$ and the parameter for the kernel calibration $\gamma= 0.01$. For the U-Net, the pre-trained weights provided by \cite{unetmri2017} are used. The weights are trained on the single-coil fastMRI knee data. The same weights are applied to out-of-distribution experiments, as described below. For GRAPPA and the fastMRI U-Net, which reconstruct the image slice by slice, we generate a 3D image by stacking the slices to a 3D volume. 

We evaluate the consistency of each method for several under-sampling operators according to specific distributions, namely Uniformly random distribution with acceleration factor $2\times$ and $15\%$ center fraction and Gaussian random distribution with acceleration factor $8\times$ and $8\%$ center fraction, respectively \cite{zbontar2018fastmri, dwork2021fast}. In order to evaluate the out-of-distribution reconstruction, we use the model that was trained on fastMRI data without any fine-tuning. The model weights provided by DiffusionMBIR \cite{chung2023solving} are taken for a fair comparison.
For fastMRI knee data.

\section{Hyper-parameters selection}
\label{App:Hyperparameters}
\begin{figure}[!thb] 
\centering
\includegraphics[width=0.8\textwidth]{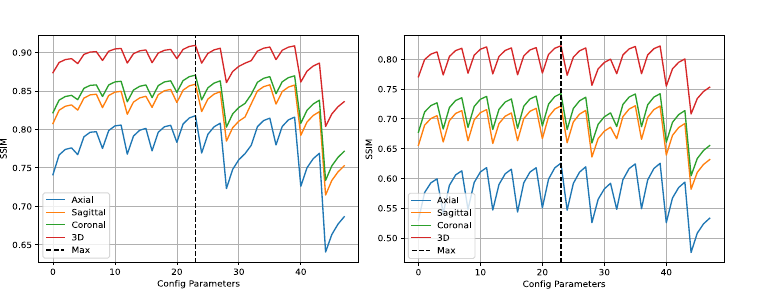}
\caption{Hyperparameters search for uniform mask (left) and Gaussian mask (right), where the maximum is given in the configuration at index = $23$ with combination parameters: $(m=10, \alpha = 0.02, \eta = 0.01, $tv=1$)$}
\label{Fig:hyperparams_app}
\end{figure}

We perform a grid search to find a good configuration for the hyper-parameters of the Fourier slice optimization, $m$, $\eta$ and $\alpha$. \Cref{Fig:hyperparams_app} shows the grid search results, where the optimal configuration is found as iteration $m=10$, regularization parameters for sparsity $\alpha = 0.02$, learning rate $\eta = 0.01$, and including total variation. We this configuration in all empirical evaluations, unless differently specified.

\section{Ablation Gaussian} \label{App:ab_gaussian}
Our results are consistent with the Uniform case, where regularization is required to improve the SSIM. This confirms that both regularization terms are contributing to improve the given optimization pipeline. 

\
\begin{figure}[!thb] 
\label{Fig:ablation_gaussian}
\centering
\includegraphics[width=0.8\textwidth]{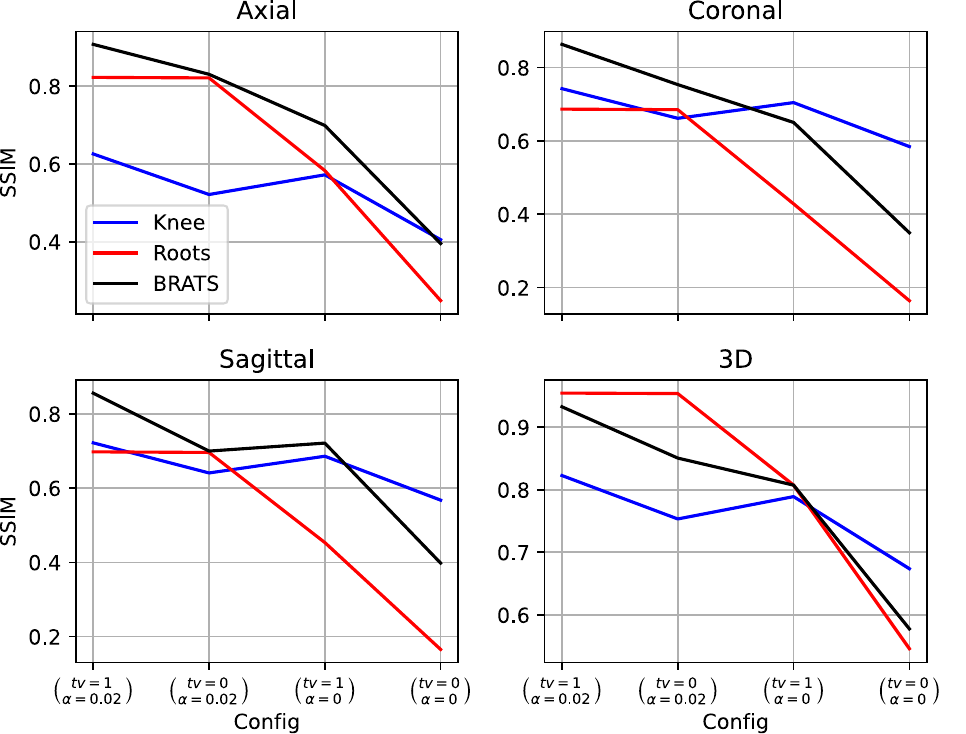}
\caption{Ablation study to investigate the effect of regularization for reconstruction in terms of SSIM $(\uparrow)$ metric for Gaussian mask. The optimization iteration $m=10$ and $\eta = 0.01$.}
\end{figure}
\section{3D MR images and their projection} \label{sec:FourierSlice}
We follow the definition and presentation of the Fourier slice method as given in \cite{bracewell1990numerical, bracewell1956strip}. Suppose we have 2D function $p(x,y)$, the projection can be written as
\begin{equation*}
p(x) = \int_{-\infty}^{\infty} p(x,y) dy
\end{equation*}
Now observe the 2D Fourier transform $\widehat{p}(k_x,k_y) = \mcl{F}_{2D} \left( p(x,y)\right)$
\begin{equation*}
\widehat{p}(k_x,k_y) ) = \int_{-\infty}^{\infty} \int_{-\infty}^{\infty} p(x,y) e^{-2\pi(x k_x + y k_y)} dx dy
\end{equation*}
Focusing on the slice at the center frequency at $y-$axis, i.e.,  $k_y = 0$, we have
\begin{equation*}
\begin{aligned}
\widehat{p}(k_x,0)  &= \int_{-\infty}^{\infty} \int_{-\infty}^{\infty} p(x,y) dy \, e^{-2\pi x k_x} dx \\&= \int_{-\infty}^{\infty} p(x) e^{-2\pi x k_x} dx = \mcl{F}_{1D} \left( p(x)\right)
\end{aligned}
\end{equation*}
This is a 1D Fourier transform on the projection. Additionally, the projection in the discrete setting is the summation on the one axis.

\section{Optimization problem details} \label{Sect:Prox_Regul}
The algorithm \ref{algo:DDPM_Sampling_Proximal} performs an alternating update between the DDPM sampling and the Fourier slice optimization in \eqref{eq:opt}. In this section, we will discuss the implementation details to solve the optimization in \eqref{eq:opt}. The sparsity constraint by using $\ell_1$-norm is non-smooth. The implementation of the regularization function in the optimization problem can be represented as the proximal operator \cite{parikh2014proximal}, namely the soft thresholding operator. Soft thresholding operator can be written as \cite[eq. 15.22]{foucart2013mathematical}.
\begin{equation}
  \label{eq: prox_soft_thres}\text{prox}_\alpha\left(\textbf{\emph X}\right)=\begin{cases}
			\frac{\textbf{\emph X}}{\card{\textbf{\emph X}}}\circ \left(\card{\textbf{\emph X}} - \alpha\right), & \text{if $\card{\textbf{\emph X}}  \geq \alpha$}\\
            0, & \text{otherwise},
		 \end{cases}
\end{equation}
where $\alpha$ is the pre-determined threshold value. It should be noted that the absolute value is applied element-wise for volumetric data. The combination of the proximal operator and the Lagrangian function is called proximal gradient method
\cite{bolte2014proximal, parikh2014proximal} described as follows
\begin{equation}
\begin{aligned}
& \underset{\textbf{\emph X} \in \C^{S \times N \times N}}{\text{minimize}}
& & \tikzmarkin[set fill color=green!50!lime!30,
set border color=green!40!black]{a-1}(0.05,-0.45)(-0.05,0.65)
\sum_{s = 1}^S \norm{\hat{\mbf{Y}}_s - \mbf{M} \circ \left( \mcl{F}_{2D}\left( \mbf{X}_s \right)\right) }_F^2
\tikzmarkend{a-1}
+ \mcl{R}\left(\textbf{\emph X} \right) \\
& \text{subject to}
& & \tikzmarkin[set fill color=orange!30,
set border color=orange!40!black]{b-1}(0.05,-0.2)(-0.05,0.37) \hat{\mbf{y}}_s^{k_y} = \mbf{m}^{k_y} \circ \mcl{F}_{1D}\left( \mcl{P}_{y} \left(\mbf{X}_s\right)\right) \quad  \text{for } s \in [S]\\
&&& \hat{\mbf{y}}_s^{k_x} =  \mbf{m}^{k_x} \circ \mcl{F}_{1D}\left( \mcl{P}_{x} \left(\mbf{X}_s\right)\right) \tikzmarkend{b-1}
\end{aligned}
\end{equation}
where the regularization $\mcl{R}\left(\textbf{\emph X}\right)$ can be written as
\begin{equation}
\begin{aligned}
  &\alpha \underbrace{\sum_{s=1}^S\sum_{i=1}^N\sum_{j=1}^N \card{x_{s,i,j}}}_{\tikzmarkin[set fill color=red!30,
set border color=red!40!black]{d-1}(0.05,-0.15)(-0.05,0.25)\text{sparsity}\tikzmarkend{d-1}} +\underbrace{\sum_{s=1}^S \sum_{i=1}^N\sum_{j=1}^{N-1}  \card{x_{s,i,j+1} - x_{s,i,j}}^2}_{\tikzmarkin[set fill color=blue!30,
set border color=blue!40!black]{c-1}(0.05,-0.15)(-0.05,0.25)\text{smoothness} \left(\mcl{TV}\left(\textbf{\emph X}\right) \right)\tikzmarkend{c-1}}  &+  \underbrace{\sum_{s=1}^S\sum_{i=1}^{N-1}\sum_{j=1}^{N} \card{x_{s,i + 1,j} - x_{s,i,j}}^2}_{\tikzmarkin[set fill color=blue!30,
set border color=blue!40!black]{c-2}(0.05,-0.15)(-0.05,0.25)\text{smoothness}\left(\mcl{TV}\left(\textbf{\emph X}\right) \right)\tikzmarkend{c-2}} 
\end{aligned}
\end{equation}
To summarize, the entire optimization problem can be formulated in \eqref{eq:proximal} and \eqref{eq:loss_function}. The proximal projection in \eqref{eq:proximal} deals with the $\ell_1-\text{norm}$, which encourages the sparsity of the reconstructed image. The first term on the right-hand-side of \eqref{eq:loss_function} ensures data fidelity on $xy-$plane, while the second and third terms enhance the continuity on $z-$axis. The last term in \eqref{eq:loss_function} incorporates smoothness of MR images.
\begin{equation}
\label{eq:proximal}
    \textbf{\emph X}^{\left(i\right)} = 
    \tikzmarkin[set fill color=red!30,
set border color=red!40!black]{d}(0.05,-0.2)(-0.05,0.35){\text{prox}_{\alpha}}\tikzmarkend{d}
\left(\textbf{\emph X}^{\left(i-1\right)} - \lambda \nabla_{\textbf{\emph X}^{\left(i-1\right)}} \widehat{L}\left(\textbf{\emph X}^{\left(i-1\right)} \right)  \right) \in \C^{S \times N \times N}
\end{equation}
where the function $ \widehat{L}\left(\textbf{\emph X}  \right)$  can be written as
\begin{equation}
\footnotesize
\label{eq:loss_function}
\begin{aligned}
\widehat{L}\left(\textbf{\emph X}, \rho \right) 
&=  \tikzmarkin[set fill color=green!50!lime!30,
set border color=green!40!black]{a}(0.05,-0.4)(-0.05,0.55)
\frac{1}{2} \sum_{s = 1}^S\norm{\hat{\mbf{Y}}_s - \mbf{M} \circ \left( \mcl{F}_{2D}\left( \mbf{X}_s \right)\right)}_F^2  
\tikzmarkend{a} + \tikzmarkin[set fill color=blue!30,
set border color=blue!40!black]{c}(0.05,-0.15)(-0.05,0.25)
\mcl{TV}\left(\textbf{\emph X}\right) 
\tikzmarkend{c}\\ 
&+ \tikzmarkin[set fill color=orange!30,
set border color=orange!40!black]{b}(0.05,-0.4)(-0.05,0.55)
\frac{1}{2}\sum_{s = 1}^S\norm{ \hat{\mbf{y}}_s^{k_y} - \mbf{m}^{k_y} \circ \mcl{F}_{1D}\left( \mcl{P}_{y} \left(\mbf{X}_s\right)\right)}_2^2 
+ \frac{1}{2}\sum_{s = 1}^S\norm{ \hat{\mbf{y}}_s^{k_x} -  \mbf{m}^{k_x} \circ \mcl{F}_{1D}\left( \mcl{P}_{x} \left(\mbf{X}_s\right)\right)}_2^2  
\tikzmarkend{b}
\end{aligned}
\end{equation}

In general the function $\mcl{G}\left(\textbf{\emph X}\right)$ in \eqref{eq:opt} can be written as in \eqref{eq:proximal}. In Appendix \ref{App:Hyperparameters}, we will discuss the choosing of hyperparameter, such as learning rate $\lambda$, proximal parameters $\alpha$, and the optimization iteration $m$, as well as the effect on the regularization function.

\section{Qualitative results}
\label{App:Qualitative}
\begin{figure*}[htb]
    \centering
    \includegraphics[width=0.8\textwidth]{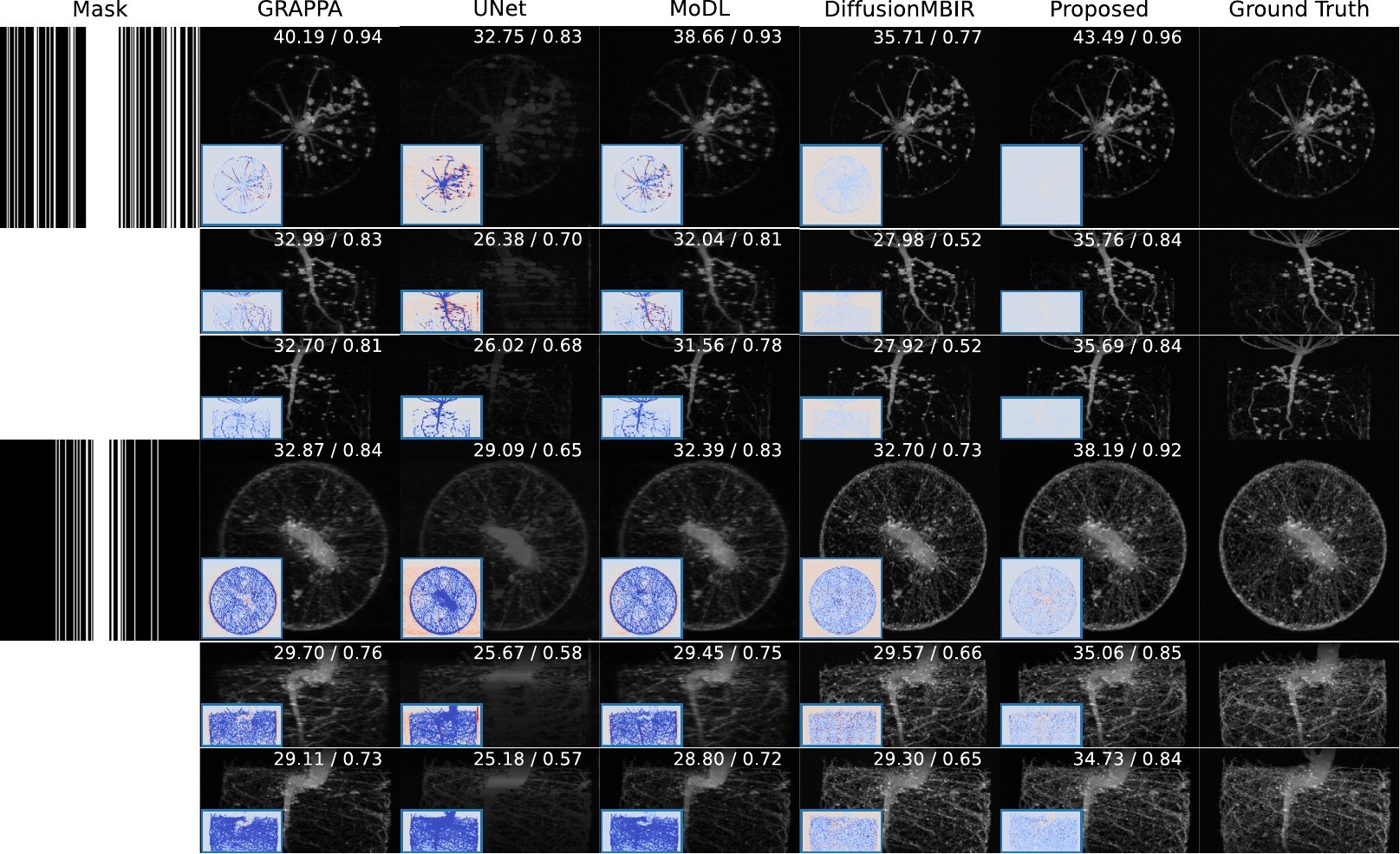}
    \caption{Reconstructions for Lotus (top) and Vicia (bottom) plant roots data.
    We present vertically the axial, coronal, and sagittal maximum intensity projections. The numbers on the upper right of each image represent the mean PSNR/SSIM of slices along dimensions. The subplots on the lower left are the difference map of the projection w.r.t. the ground truth. The color range is between $-0.1$ (bluish) and $0.1$ (reddish).}
\label{Fig:lotus_projection}
\end{figure*}

In \Cref{Fig:lotus_projection} the maximum intensity projections on each axis are displayed to visualize the fine roots structure. In addition, we report the average SSIM and PSNR for each view (axial, coronal, sagittal). It is evident that the fastMRI U-Net struggles to reconstruct the image in areas with limited information, regardless of Uniform or Gaussian masking in the column direction. Besides achieving the highest SSIM and PSNR, the proposed method generates a high-contrast image with minimal pixel-wise differences compared to other methods, as depicted in the bottom-left subplots.

\begin{figure*}[htb]
    \centering
    \includegraphics[width=0.8\textwidth]{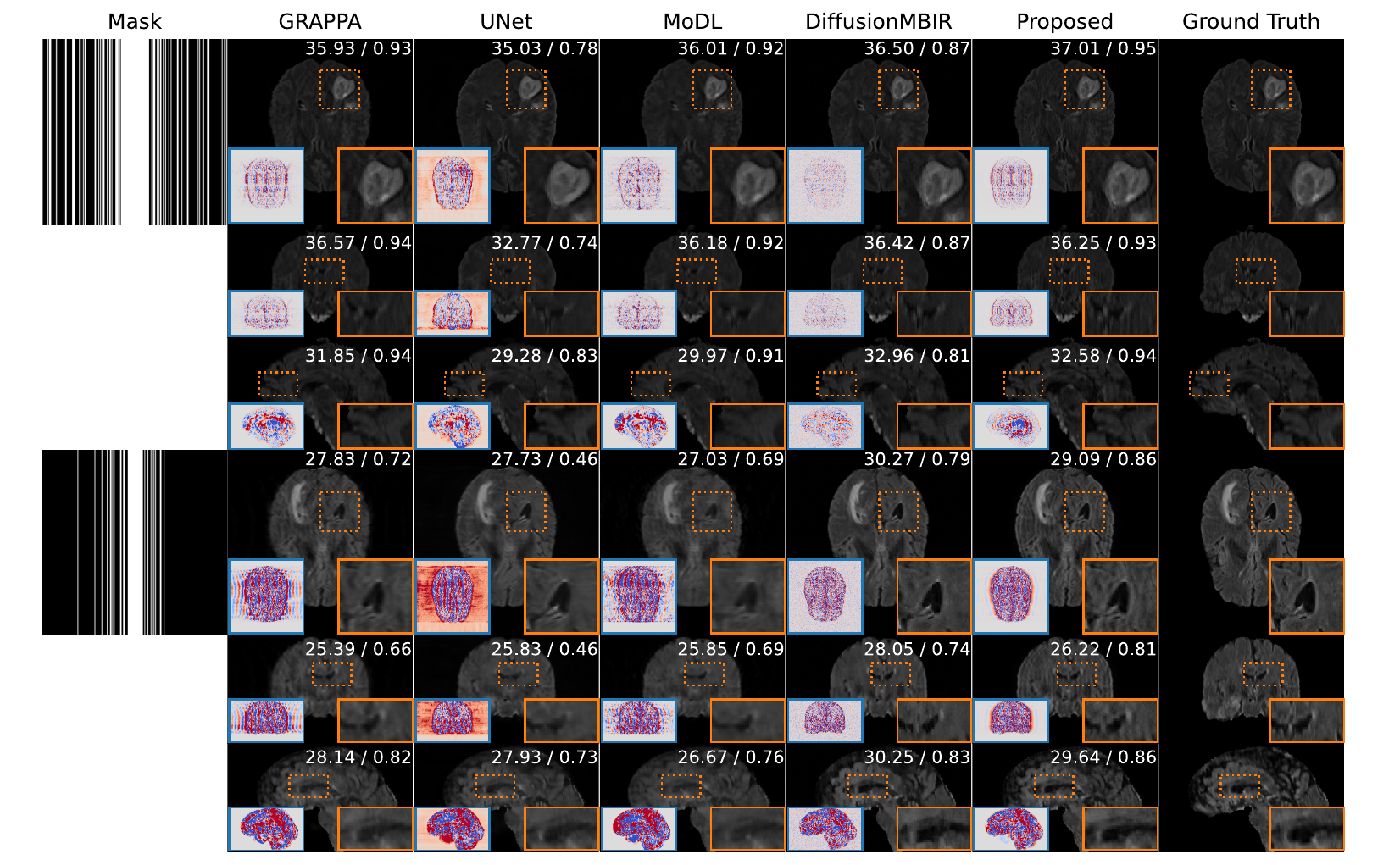}
    \caption{Slices from the volume reconstructions for BRATS data from Brats18\_CBICA\_APM\_1 (top) and Brats18\_CBICA\_AAM\_1 (bottom). 
    We present vertically the axial, sagittal, and coronal middle slices. The numbers on the upper right of each image represent the PSNR/SSIM of middle slices. The subplots on the lower left are the difference map of the projection w.r.t. the ground truth. The color range is between $-0.02$ (bluish) and $0.02$ (reddish).}
\label{Fig:brats_slice}
\end{figure*}

\Cref{Fig:brats_slice} illustrates the reconstruction of BRATS data for each view, accompanied by SSIM and PSNR values for each slice. The figure highlights that DiffusionMBIR yields smaller deviations in error differences with the ground truth, as depicted in the lower left corner box. However, it also introduces higher background noise compared to the proposed method. 
Notably, in the zoomed area with Gaussian masking, the proposed method effectively generates finer details while adhering closely to the ground truth in comparison to its counterparts.

\end{document}